%
\input harvmac
 
\font\ticp=cmcsc10
 
\def\Title#1#2{\rightline{#1}\ifx\answ\bigans\nopagenumbers\pageno0\vskip1in
\else\pageno1\vskip.8in\fi \centerline{\titlefont #2}\vskip .5in}

\font\ticp=cmcsc10
\font\ttsmall=cmtt10 at 8pt

\input epsf
\ifx\epsfbox\UnDeFiNeD\message{(NO epsf.tex, FIGURES WILL BE
IGNORED)}
\def\figin#1{\vskip2in}
\else\message{(FIGURES WILL BE INCLUDED)}\def\figin#1{#1}\fi
\def\ifig#1#2#3{\xdef#1{fig.~\the\figno}
\goodbreak\topinsert\figin{\centerline{#3}}%
\smallskip\centerline{\vbox{\baselineskip12pt
\advance\hsize by -1truein\noindent{\bf Fig.~\the\figno:} #2}}
\bigskip\endinsert\global\advance\figno by1}

\writedefs
%
%
\def\lam{\lambda}

\def\[{\left [}
\def\]{\right ]}
\def\({\left (}
\def\){\right )}


\lref\WangGS{
B.~Wang, C.~Y.~Lin and E.~Abdalla,
Phys.\ Lett.\ B {\bf 481}, 79 (2000)
[arXiv:hep-th/0003295].
}
\lref\ReyBQ{
S.~J.~Rey, S.~Theisen and J.~T.~Yee,
``Wilson-Polyakov loop at finite temperature in large N gauge theory and  anti-de Sitter supergravity,''
Nucl.\ Phys.\ B {\bf 527}, 171 (1998)
[arXiv:hep-th/9803135].
}
\lref\BrandhuberBS{
A.~Brandhuber, N.~Itzhaki, J.~Sonnenschein and S.~Yankielowicz,
``Wilson loops in the large N limit at finite temperature,''
Phys.\ Lett.\ B {\bf 434}, 36 (1998)
[arXiv:hep-th/9803137].
}
\lref\PolicastroYC{
G.~Policastro, D.~T.~Son and A.~O.~Starinets,
``The shear viscosity of strongly coupled N = 4 supersymmetric Yang-Mills  plasma,''
Phys.\ Rev.\ Lett.\  {\bf 87}, 081601 (2001)
[arXiv:hep-th/0104066].
}

\lref\GubserDE{
S.~S.~Gubser, I.~R.~Klebanov and A.~W.~Peet,
``Entropy and Temperature of Black 3-Branes,''
Phys.\ Rev.\ D {\bf 54}, 3915 (1996)
[arXiv:hep-th/9602135].
}
\lref\GubserNZ{
S.~S.~Gubser, I.~R.~Klebanov and A.~A.~Tseytlin,
``Coupling constant dependence in the thermodynamics of N = 4  supersymmetric Yang-Mills theory,''
Nucl.\ Phys.\ B {\bf 534}, 202 (1998)
[arXiv:hep-th/9805156].
}
\lref\chandra{S.~Chandrasekhar, {\it 
``The Mathematical Theory of Black Holes''}, 
Oxford University Press, New York, 1983.} 

\lref\deMelloKochQS{
R.~de Mello Koch, A.~Jevicki, M.~Mihailescu and J.~P.~Nunes,
``Evaluation of glueball masses from supergravity,''
Phys.\ Rev.\ D {\bf 58}, 105009 (1998)
[arXiv:hep-th/9806125].
}
\lref\CsakiQR{
C.~Csaki, H.~Ooguri, Y.~Oz and J.~Terning,
``Glueball mass spectrum from supergravity,''
JHEP {\bf 9901}, 017 (1999)
[arXiv:hep-th/9806021].
}
\lref\ZyskinTG{
M.~Zyskin,
Phys.\ Lett.\ B {\bf 439}, 373 (1998)
[arXiv:hep-th/9806128].
}
\lref\MinahanTM{
J.~A.~Minahan,
JHEP {\bf 9901}, 020 (1999)
[arXiv:hep-th/9811156].
}
\lref\CardosoHN{
V.~Cardoso and J.~P.~Lemos,
``Scalar, electromagnetic and Weyl perturbations of BTZ black holes:  Quasi normal modes,''
Phys.\ Rev.\ D {\bf 63}, 124015 (2001)
[arXiv:gr-qc/0101052].
}

\lref\CardosoBB{
V.~Cardoso and J.~P.~Lemos,
``Quasi-normal modes of Schwarzschild anti-de Sitter black holes:  Electromagnetic and gravitational perturbations,''
Phys.\ Rev.\ D {\bf 64}, 084017 (2001)
[arXiv:gr-qc/0105103].
}
\lref\MossGA{
I.~G.~Moss and J.~P.~Norman,
Class.\ Quant.\ Grav.\  {\bf 19}, 2323 (2002)
[arXiv:gr-qc/0201016].
}

\lref\KonoplyaZU{
R.~A.~Konoplya,
``On quasinormal modes of small Schwarzschild-anti-de-Sitter black hole,''
arXiv:hep-th/0205142.
}

\lref\KimEV{
W.~T.~Kim and J.~J.~Oh,
Phys.\ Lett.\ B {\bf 514}, 155 (2001)
[arXiv:hep-th/0105112].
}

\lref\BirminghamHC{
D.~Birmingham,
Phys.\ Rev.\ D {\bf 64}, 064024 (2001)
[arXiv:hep-th/0101194].
}

\lref\CardosoVS{
V.~Cardoso and J.~P.~Lemos,
``Quasi-normal modes of toroidal, cylindrical and planar black holes in  anti-de Sitter spacetimes,''
Class.\ Quant.\ Grav.\  {\bf 18}, 5257 (2001)
[arXiv:gr-qc/0107098].
}
\lref\HorowitzJD{
G.~T.~Horowitz and V.~E.~Hubeny,
``Quasinormal modes of AdS black holes and the approach to thermal  equilibrium,''
Phys.\ Rev.\ D {\bf 62}, 024027 (2000)
[arXiv:hep-th/9909056].
}
\lref\MaldacenaRE{
J.~M.~Maldacena,
``The large $N$ limit of superconformal field theories and supergravity,''
Adv.\ Theor.\ Math.\ Phys.\  {\bf 2}, 231 (1998)
[Int.\ J.\ Theor.\ Phys.\  {\bf 38}, 1113 (1999)]
[arXiv:hep-th/9711200].
}
\lref\GubserBC{
S.~S.~Gubser, I.~R.~Klebanov and A.~M.~Polyakov,
``Gauge theory correlators from non-critical string theory,''
Phys.\ Lett.\ B {\bf 428}, 105 (1998)
[arXiv:hep-th/9802109].
}
\lref\WittenQJ{
E.~Witten,
``Anti-de Sitter space and holography,''
Adv.\ Theor.\ Math.\ Phys.\  {\bf 2}, 253 (1998)
[arXiv:hep-th/9802150].
}
\lref\leaver{E.W.~Leaver, ``An analytic representation for the quasi-normal modes of Kerr black holes'', Proc. \ R. \ Soc. \ Lond. \ A {\bf 402} 285 (1985).}
\lref\jaffe{G.~Jaff\'{e}, ``Zur Theorie des Wasserstoffmolek\H{u}lions'', 
Z.\ Phys. \ {\bf 87}, 535 (1933).}
\lref\BirminghamPJ{
D.~Birmingham, I.~Sachs and S.~N.~Solodukhin,
``Conformal field theory interpretation of black hole quasi-normal modes,''
Phys.\ Rev.\ Lett.\  {\bf 88}, 151301 (2002)
[arXiv:hep-th/0112055].
}
\lref\PolicastroSE{
G.~Policastro, D.~T.~Son and A.~O.~Starinets,
``From AdS/CFT correspondence to hydrodynamics I,II,''
arXiv:hep-th/0205052, hep-th/0210220.
}

\lref\SonSD{
D.~T.~Son and A.~O.~Starinets,
``Minkowski-space correlators in AdS/CFT correspondence: Recipe and  applications,''
JHEP {\bf 0209}, 042 (2002)
[arXiv:hep-th/0205051].
}
\lref\nollert{H.-P. Nollert, ``Quasinormal modes of Schwarzschild black holes: The determination of quasinormal frequencies with very large imaginary parts'', Phys.\ Rev. D47 (1993) 5253.}
\lref\askey{{\it Theory and Application of Special Functions}, Ed. R.~Askey,
Academic Press, New York (1975).}
\lref\pincherle{S.~Pincherle, ``Delle funzioni ipergeometriche e de varie questione ad esse attinenti'', Giorn.\ mat.\ Battaglini {\bf 32} (1894) 209.}
\lref\schafke{R.~Sch\"{a}fke and D.~Schmidt, ``The connection problem for general linear ordinary differential equations at two regular singular points with applications to the theory of special functions'', SIAM J.\ Math.\ Anal.\ {\bf 11} (1980) 848.}
\lref\watson{E.T.~Whittaker and G.N.~Watson, 
{\it A course of modern analysis}, Cambridge University Press, 1927.} 
\lref\wongli{R.~Wong and H.~Li, ``Asymptotic expansions 
for second order linear difference equations'', J.\ Comput. 
\ Appl.\  Math.\ {\bf 41} (1992) 65.}
\lref\heunpaper{
K.~Heun, ``Zur Theorie der Riemann's sehen Functionen zweiter Ordnung 
mit vier Verzweigungspunkten'',  Math.\ Ann. {\bf 33} (1889) 61.}
\lref\bateman{
{\it Higher transcendental functions, vol.\ III}, 
ed. A.~Erdelyi,
McGraw Hill, New York (1955).}
\lref\heunbook{
{\it Heun's differential equation}, ed. A.~Ronveaux, F.~Arscott,
Oxford University Press, Oxford (1995).}

%
%
\baselineskip 16pt
\Title{\vbox{\baselineskip12pt
\line{\hfil   INT-PUB-02-44}
\line{\hfil \tt hep-th/0207133} }}
{\vbox{
{\centerline{Quasinormal Modes of Near Extremal Black Branes }}
}}
\centerline{\ticp Andrei O. Starinets\footnote{}{\ttsmall
 starina@phys.washington.edu}}
\bigskip
\centerline{\it Institute for Nuclear Theory, University of Washington, Seattle, WA 98195}
\bigskip
\centerline{\bf Abstract}
\bigskip
We find quasinormal modes of near extremal black branes by 
solving a singular boundary value problem for the Heun equation.
The corresponding eigenvalues determine the dispersion law 
for the collective 
excitations in the dual strongly coupled 
${\cal N}=4$ supersymmetric Yang-Mills theory at finite temperature.

\bigskip

$\,$

\bigskip

\centerline{\sl In memory of Aleksandr Sergeevich Vshivtsev}

\Date{July 2002}
\newsec{Introduction}

The gauge theory/gravity correspondence \refs{\MaldacenaRE ,\GubserBC , 
\WittenQJ} 
provides useful insights into the 
properties of a strongly coupled supersymmetric Yang-Mills (SYM) theory at 
non-zero temperature. A well-known example is the finite-temperature 
${\cal N}=4$ $SU(N)$ 
SYM theory in $4d$
which in the large $N$, large 't Hooft coupling limit is 
dual to the gravitational background of $N$ parallel near extremal 
three-branes, with temperature related to the parameter of 
non extremality. For this theory, a 
number of quantities such as the free energy \refs{\GubserDE ,\GubserNZ},
the Wilson loop \refs{\ReyBQ , \BrandhuberBS},
the shear viscosity \refs{\PolicastroYC ,\PolicastroSE}, 
the $R$-charge diffusion constant \PolicastroSE\ , and the 
Chern-Simons diffusion rate \SonSD\ have been computed using the methods of 
 gauge theory/gravity duality. Since  supersymmetry is broken 
at finite temperature and thus no non-renormalization theorem is expected to 
hold, quantities computed in the strong coupling regime using a
gravity dual differ from their analogues obtained at weak coupling 
via perturbation theory. The results of a gravity calculation are then 
regarded as a prediction for the SYM theory, 
assuming that the AdS/CFT correspondence 
is valid at finite temperature\foot{A non-trivial check of the validity of
 the correspondence at finite temperature 
was made recently in \PolicastroSE\ .}. 

Dynamical properties of a thermal gauge theory are encoded in its 
Green's functions. In the context of AdS/CFT, 
Minkowski space Green's functions can be computed from
gravity using the recipe given in \SonSD . Unfortunately, 
for a non-extremal background,
only approximate expressions for the correlators are usually obtained.
For example, the retarded propagator of the gauge invariant local 
operator $F^2$ (dual to the dilaton) defined by
\eqn\prop{ G^R(\omega, {\bf k}) = -\, i \int d t\, d^3 x \,
e^{-i\omega t + i {\bf k} {\bf x}}\, \theta (t)\, \langle [ F^2(x),
F^2(0)]\rangle }
can be explicitly computed only at zero or very high (with respect to 
the absolute value of momentum) temperature. 
At zero temperature, the retarded propagator \prop\ has a branch cut 
singularity for $|\omega| >  |{\bf k}|$,
\eqn\zerot{ G^R(\omega, {\bf k})  = {N^2 (-\omega^2 + {\bf k}^2)^2
\over 64 \pi^2}
 \left( \ln{ |-\omega^2 + {\bf k}^2|} - i \pi 
\theta (\omega^2 - {\bf k}^2)\, \hbox{sgn}\, \omega \right)\,.}
In the high temperature limit $\omega/T \ll 1$, $|{\bf k}|/T \ll 1$, 
the propagator is analytic in the complex 
 $\omega$-plane,
\eqn\hight{ G^R(\omega, {\bf k})  = - { N^2 T^2\over 16}\left( i\,2 
 \pi T \omega 
+  {\bf k}^2 \right)\,.}
However, for 
  generic values of $\omega$ and ${\bf k}$, we expect 
$G^R(\omega, {\bf k})$
to have poles corresponding to the 
spectrum of collective excitations of the SYM plasma. 

One can compare the situation to the simpler case of the $2d$ CFT dual to the 
 Ba\~nados-Teitelboim-Zanelli (BTZ) black hole background. There, the 
 retarded Green's functions can be computed exactly.  
For illustration, consider the 
case of the conformal dimension $\Delta =2$. Then
\eqn\btz{G^R_{2d} (\omega,k) = {\omega^2-k^2\over 4\pi^2}\left[
\psi \left( 1 - i\, {\omega - k\over 4\pi T} \right) +
\psi \left( 1 - i\, {\omega + k\over 4\pi T} \right)
\right]\,,}
where we have put $T_L=T_R$ and ignored the constant prefactor for simplicity.
The high temperature limit of \btz\ is an analytic function of $\omega$.
In general, however, $G^R_{2d} (\omega,k)$ has infinitely many poles located 
at 
\eqn\btzpoles{\omega_n = \pm \, k - i\, 4\pi \, T \, (n+1)\,, \;\;\;\;\; n=0,1,\dots \,.}
When $T\rightarrow 0$, the poles merge forming branch cuts. One can
use the asymptotic expansion $\psi (z) \sim \log{z} - 1/2z +\dots$ (valid
for $|\hbox{arg}\,  z| < \pi/2$) to find the zero-temperature limit of \btz\
(ignoring the contact terms) :
\eqn\rebtz{ G^R_{2d} \sim {\omega^2-k^2\over 4\pi^2} 
\log{|\omega^2-k^2|} - i\, {\omega^2-k^2\over 4\pi}\theta (\omega^2-k^2)
\; \hbox{sgn}\,\, \omega\,.}
For spacelike momenta, $|\omega|<k$, the imaginary part of $G^R_{2d}$ is 
exponentially suppressed.

Even though the retarded Green's function in $4d$ cannot be found explicitly,
the location of its singularities and thus the ``dispersion law'' of 
thermal excitations at strong coupling can be determined precisely. 
As shown in \SonSD\ , this amounts to finding the quasinormal 
frequencies of dilaton's fluctuation in 
the dual near-extremal black brane background as 
functions of the spatial momentum\foot{Note that the poles \btzpoles\ of 
 $G^R_{2d} (\omega,k)$ coincide with the quasinormal frequencies of
BTZ black hole \BirminghamPJ .} 
. Recently, there has been considerable interest (inspired by the 
AdS/CFT duality conjecture) in studying quasinormal modes of
Schwarzschild- AdS black holes\foot{References to
the early works on  quasinormal modes in asymptotically AdS 
spacetime as well as to works considering bulk dimension other 
than five can be found in  \SonSD\ , \HorowitzJD .}
 \refs{\HorowitzJD , \WangGS , \CardosoBB ,  
\KimEV , \CardosoVS , \MossGA , \KonoplyaZU}. 
Here, we consider the non-compact case directly. This corresponds 
to finding  quasinormal frequencies for an infinitely large
AdS black hole.

Computing the scalar quasinormal frequencies in the 
black brane background is equivalent 
 to solving the two-parameter singular spectral 
problem for the equation with four regular singularities (Heun equation).
In this paper, we solve this problem by using the elegant method based on 
Pincherle's theorem. The eigenvalue equation is written in terms of
continued fractions and solved numerically. 
Similar approach had been used by G.~Jaff\'{e} \jaffe\ in 1933 to 
find the spectrum of the hydrogen molecule ion. Later, it has been applied
to the problem of quasinormal modes in asymptotically flat spacetimes
by E.W.~Leaver \leaver\ .

The paper is organized as follows. In the next section, we formulate the 
problem of finding the quasinormal modes of near-extremal black three-brane 
in terms of the singular boundary value problem for Heun equation.
In section 3 this boundary value problem is solved by analyzing the 
associated linear difference equation and applying Pincherle's theorem.
The results (quasinormal frequencies $\omega_n$ and the ``dispersion law''
$\omega ({\bf k})$)
are presented in section 4. The discussion follows in section 5.

\newsec{Quasinormal modes and the boundary value problem for Heun equation}

The metric corresponding to the collection of $N$ parallel 
non-extremal three-branes in the near-horizon limit is given by
\eqn\metric{ds^2 = {r^2\over R^2} \left( - f dt^2 + d{\bf x}^2\right)
+  {R^2\over r^2}\left( f^{-1} dr^2 + r^2 d\Omega_5^2\right)\,, }
where $f(r) = 1 - r_0^4/r^4$. According to the gauge 
theory/gravity correspondence, 
this background with the 
parameter of non-extremality $r_0$ is dual to the ${\cal N}=4$ 
$SU(N)$ SYM at finite temperature $T=r_0/\pi R^2$ in the limit
$N\rightarrow \infty$, $g^2_{YM} N\rightarrow \infty$.

Using the new coordinate $z=1-r_0^2/r^2$ and the Fourier decomposition
\eqn\fourier{\phi(z,t,{\bf x}) = \int \! {d^4 k\over (2\pi)^4 }
e^{-i\omega t + i {\bf k}\cdot{\bf x}}\phi_k(z)\,, }
 the equation for the 
minimally coupled massless scalar in the background \metric\ reads
\eqn\scal{\phi_k'' + {1+(1-z)^2\over z (1-z)(2-z)}\, \phi_k' +
{\lambda^2\over 4 z^2(1-z)(2-z)^2}\, \phi_k 
- {q^2\over 4 z (1-z)(2-z)} \, \phi_k = 0\,,}
where $\lambda = \omega/\pi T$, $q = |\vec{k}|/\pi T$.
Eq. \scal\ has four regular singularities at $z=0,1,2,\infty$,
the corresponding pairs of characteristic exponents being 
respectively $\{-i\lambda/4, i\lambda/4 \}$; 
$\{0,2\}$; $\{-\lambda/4,\lambda/4\}$; $\{0,0\}$.

Quasinormal modes are defined as solutions of Eq.\scal\ obeying the
``incoming wave'' boundary condition at the horizon $z=0$ and the
vanishing Dirichlet boundary condition at spatial infinity $z=1$.
The first condition singles out the exponent $\nu_0^{(1)}
 = -i\lambda/4$ at $z=0$.

The most straightforward way to find quasinormal modes would be to
construct a local series solution $\varphi_{loc}(z,\lambda)$ to Eq.\scal\ 
with the exponent $\nu_0^{(1)}$ near the origin, prove its convergence 
at $z=1$, and determine the eigenfrequencies $\lambda_n (q)$ 
by solving  the equation
$\varphi_{loc}(1,\lambda)=0$ numerically. This approach
works quite well for the low-level eigenfrequencies, 
and in fact it has been  
successfully used in a number of publications on 
quasinormal modes in asymptotically $AdS$ space.
Here we would like to solve the above 
eigenvalue problem in a somewhat different way which, in our opinion,
is more appealing both analytically and numerically.

By making a transformation of the dependent variable
\eqn\change{ \phi (z) = 
z^{-{i\lambda\over 4}}\,(z-2)^{-{\lambda\over 4}}\, y(z)\,,}
Eq.\scal\ can be reduced to the standard form of  the Heun 
equation \refs{\heunpaper,\heunbook},
\eqn\heun{ y'' + \left[ {\gamma\over z} +{\delta\over z-1} + 
{\epsilon\over z-2}\right] y' + {\alpha\beta z - Q\over z(z-1)(z-2)}y = 0\,,}
where $\alpha$, $\beta$, $\gamma$, $\epsilon$ depend on $\lambda$,
\eqn\albet{\alpha = \beta = -{\lambda (1+i)\over 4}\,,
\;\;\; \;\; \gamma = 1 -{i\lambda\over 2}\,,\;\;\;\; \delta = -1\,, 
\;\;\;\; \epsilon = 1-\lambda /2\,, }
and $Q$ is the so called ``accessory parameter'' given in our case by
\eqn\Q{ Q = {q^2\over 4} - {\lambda (1-i)\over 4} - {\lambda^2(2-i)\over 8}\,.}
We would like to determine values of $\lambda$ and $q$ for which 
 Eq.~\heun\  on the interval $[0,1]$ 
has solutions obeying the boundary conditions $y(0)=1$, $y(1)=0$.

Before turning to the solution, let us remark that Eq.~\scal\ with 
$\lam =0$ has received some attention previously in  connection 
with the so called ``glueball mass spectrum'' in $QCD_3$ \CsakiQR ,
\deMelloKochQS ,\ZyskinTG , \MinahanTM . 
Approximate analytic expressions for 
the ``glueball'' massess squared $M^2_n = - q^2_n$ were found 
either via the ``brute force'' WKB calculation \MinahanTM\
or by appealing to the 
unpublished results of the RIMS group  \CsakiQR\  
(both approaches give satisfactory agreement with the 
results obtained by numerical integration). Introducing
a non-zero $\lambda$ and using the ``incoming wave'' boundary condition
makes the problem considerably more complicated, as is evident from 
Eqs.~\heun\ and
\albet\ .

\newsec{Solving the boundary value problem}

\subsec{Local solutions}
The Frobenius set of local solutions near each of the
 singularities can be easily constructed.

At $z=0$, the local series solution corresponding to 
the index $\nu =0$ and normalized to 1 is given by
\eqn\frobeniuszero{y_0 (z) = \sum\limits_{n=0}^{\infty} a_n (\lambda,k)\, z^n\,,}
where $a_0=1$, $a_1 = {Q\over 2\gamma}$, and the coefficients $a_n$ with
$n\geq 2$ obey the three-term recursion relation
\eqn\recurrence{a_{n+2} + A_n(\lam) \, a_{n+1} + B_n(\lam) \, a_n = 0\,, }
where
\eqn\adef{A_n(\lam) = - {(n+1)(2\delta +\epsilon + 3(n+\gamma ))
+ Q\over 2(n+2)(n+1+\gamma)}\,,}
\eqn\bdef{B_n(\lam) =  {(n+\alpha)(n+\beta)\over 2(n+2)(n+1+\gamma)}\,.}
The series \frobeniuszero\ is absolutely convergent for $|z|<1$ and, 
in general, is divergent for $|z|>1$. 
The condition for convergence at $|z|=1$ involves parameters of 
the equation and will be investigated below.

At $z=1$, the difference of the exponents has an integer
value, and we expect the local solution there to contain logarithms.
Indeed, the set of local solutions is given by
\eqn\froboneone{y_1(z) = (1-z)^2 \left( 1 + b_1^{(1)} (1-z) + 
b_2^{(1)} (1-z)^2 +\cdots\right)\,,} 
\eqn\frobonetwo{y_2(z) = 1 + b_1^{(2)} (1-z) + h y_1(z) \log{(1-z)}
+ b_2^{(2)} (1-z)^2 +\cdots \,,}
where $b_1^{(1)} = (q^2 - \lambda^2 + 3\lambda (1-i))/12$,
 $b_1^{(2)} = (- q^2 + \lambda^2 + \lambda (1-i))/4$,
$h = - (q^2-\lambda^2)^2/32$,
and the coefficients $b_n^{(1,2)}$,
$n\geq 2$, can be found recursively from the 
relation similar to the one in \recurrence\ . Solving the differential 
equation Eq.~\heun\ essentially means finding the connection between the 
sets of local
solutions.

\subsec{The connection problem}
\subseclab\connect
From the general theory of linear differential equations, it follows that
three solutions $y_0 (z)$, $y_1(z)$, $y_2(z)$ are connected on $[0,1]$ 
by the linear relation 
\eqn\connect{ y_0(z) = {\cal A}(\lambda,q) \, y_1(z) + {\cal B}(\lambda,q)
 \, y_2(z)\,,}
where ${\cal A}$, ${\cal B}$ are independent of $z$. The 
vanishing Dirichlet boundary condition for quasinormal modes at $z=1$ implies 
 ${\cal B}=0$, which is the equation for eigenfrequencies $\lambda_n$.
If $\lambda = \lambda_n$, $y_0 (z)$ is proportional to 
 $y_1(z)$, and thus it is simultaneously a Frobenius solution 
about $z=0$ (with exponent $0$) and a Frobenius solution about $z=1$
(with exponent $2$). Such a solution is called a Heun function \bateman\ .
We learn that quasinormal modes of black branes are Heun functions. 
Unfortunately, the connection problem for the Heun equation remains unsolved,
and  explicit expressions for the coefficients ${\cal A}$, 
${\cal B}$ are unavailable, 
with the exception of some special cases\foot{Sch\"{a}fke and Schmidt
\schafke\ give the connection coefficients in terms of the 
$n\rightarrow \infty$ limit of $a_n$ obeying Eq. \recurrence\ . Although 
 the asymptotic
behavior of $a_n$ near $n=\infty$ can be studied 
by means of local analysis, it 
only determines $a_n$ up to a $\lambda$-dependent coefficient which 
is essentially the very quantity we are looking for.}. 
There is, however, an indirect way of determining for which values of
$\lambda$ and $q$ the connection coefficient  ${\cal B}$ vanishes. This is 
achieved through an analysis of convergence.

\subsec{The analysis of convergence}
\subseclab\convergence
\writedefs
The convergence of the series \frobeniuszero\ can be analyzed by 
studying the large $n$ asymptotic behavior of the 
linear difference equation \recurrence\ .
One finds\foot{Necessary information about second order linear 
difference equations can be found, for example, 
in the excellent review paper by Wong and Li \wongli\ .} 
that Eq. \recurrence\ possesses two linearly independent 
asymptotic solutions of the form
\eqn\birkone{a_n^{(1)} \sim 2^{-n} n^{-1-{\lambda\over 2}}
\sum\limits_{s=0}^{\infty} {c_s^{(1)}\over n^s}\,,}
\eqn\birktwo{a_n^{(2)} \sim  n^{-3}
\sum\limits_{s=0}^{\infty} {c_s^{(2)}\over n^s}\,,}
where the coefficients $c_s^{(1,2)}$ can be found recursively using 
the asymptotic expansion of $A(n)$, $B(n)$. In particular, we have 
$c_1^{(1)} = (q^2-\lambda^2 + \lambda (1+i)
 + 2 \lambda^2 (1+i))/4$, 
$c_1^{(2)} = (12 -  q^2 + \lambda^2 - 3 (1+i)\lambda )/4$.
Equation \birkone\  is called a {\it minimal} solution to Eq.~\recurrence\ ,
and Eq.~\birktwo\ represents a {\it dominant} one.
This distinction reflects the property 
\eqn\mindom{\lim_{n\rightarrow \infty} { a_n^{(1)}\over a_n^{(2)}} =0\,,}
and will be useful later on.
Using Eqs.~\birkone\ , \birktwo\ , we obtain 
\eqn\limsone{\left| { a_{n+1}^{(1)} \over a_n^{(1)}}\right|
 =  {1\over 2} \left[ 1 - {2 +  \hbox{Re}\, \lambda\over 2 n} + 
O\left( {1\over n^2}\right)\right]\,, }
\eqn\limstwo{ \left| {a_{n+1}^{(2)}\over a_n^{(2)}}\right| 
=  1 - {3\over n} + O \left( {1\over n^2}\right)\,.}
Equations \limsone\ , \limstwo\ imply\foot{See \watson\ , \S 2.37. }
  that generically 
the series \frobeniuszero\ converges
absolutely for $|z|\leq 1$ for {\it any} value of $\lambda$.
This proves that the equation 
\eqn\dir{\sum\limits_{n=0}^{\infty} a_n (\lambda,k) = 0\,}
can indeed be used to compute the quasinormal frequencies numerically. 
What is more interesting, however, is that in some cases the 
radius of convergence increases. It happens when the minimal 
solution \birkone\ exists. In that case the series \frobeniuszero\
converges absolutely for $|z|<2$ and thus represents an analytic function
(the Heun function) in that region. 

Thus, even though the connection coefficient ${\cal B}(\lambda,q)$ in 
Eq.~\connect\  remains unknown,
finding zeros of  ${\cal B}(\lambda,q)$  
is equivalent to finding a condition under which the minimal solution 
to Eq.~\recurrence\ exists.
Such a condition is conveniently supplied by Pincherle's theorem
\pincherle\ which states that the minimal solution exists if and only if
the continued fraction
\eqn\fraction{- {B_0(\lam)\over A_0(\lam)-}{B_1(\lam)\over A_1(\lam)-}
{B_2(\lam)\over A_2(\lam)-}\cdots }
converges. Moreover, in case of convergence one has
\eqn\fractionn{{a_{n+1}\over a_{n}} = - {B_{n}(\lam)\over A_{n}(\lam)
-}{B_{n+1}(\lam)\over A_{n+1}(\lam)-}{B_{n+2}(\lam)\over A_{n+2}(\lam)
-}\cdots \,. }
The right hand side of Eq.~\fractionn\ is generated by a simple algorithm.
Define $r_n = a_{n+1}/a_n$. Then Eq.\recurrence\ can be written as
\eqn\algo{r_n = - {B_n(\lam)\over A_n(\lam) + r_{n+1}}\,. }
Applying \algo\ repeatedly, one gets Eq.~\fractionn\ . Now, 
setting $n=0$ in Eq.~\fractionn\  we have
\eqn\eigen{{Q\over 2 - i\lam} = - 
{B_0(\lam)\over A_0(\lam)-}{B_1(\lam)\over A_1(\lam)-}
{B_2(\lam)\over A_2(\lam)-}\cdots\,. }
This is the transcendental eigenvalue equation which determines 
the quasinormal frequencies. Equation \eigen\ can be solved numerically 
with great efficiency using a variety of methods (see 
\askey\ , which also includes a discussion of the error analysis). Here 
we use a nonlinear backward recursion which amounts to breaking 
the continued fraction \algo\ by setting $r_n$ to 
zero\foot{To improve the convergence of continued fractions, 
one can instead set $r_n$ to its 
asymptotic value at $n_*$, $r_n = 1/2 - (2+\lam)/4 n_* +\cdots$,
as suggested by Nollert \nollert\ .} 
for some large $n_*$ and computing backward to get $r_0$. 
Stability can be checked by choosing a larger value of $n_*$ 
and repeating the calculation.

\newsec{Quasinormal frequencies}

For $q=0$, the lowest 15  quasinormal 
frequencies $\lam_n$ obtained by solving Eq.~\eigen\ numerically 
are listed in Table 1 and shown in 
Fig.~1. The results suggest that the number of frequencies is infinite,
and that their large $n$ asymptotic behavior is given by the simple formula
\eqn\conj{\lam_n^{\pm} = \lam_0^{\pm} \pm 2 n (1 \mp i) \,,}
where $\lam_0^{\pm} \approx \pm 
1.2139 - 0.7775\, i$. In terms of the original variable 
$\omega$, Eq.~\conj\ is written as
\eqn\conjom{\omega_n^{\pm} = \omega_0^{\pm} \pm 2 \pi T n (1 \mp i)\,,}
where the coefficient in front of the parentheses  
is the bosonic Matsubara frequency and $\omega_0^{\pm}
 = \pi T \lambda_0^{\pm}$.

%
%
\midinsert
\centerline{%
\vbox{
  \offinterlineskip \tabskip=0pt
  \halign{\strut
          \vrule#&              %
          \hfil $ #~$ &\vrule#& %
          \hfil $\,#$ &         %
        ~ \hfil $#$ &\vrule#&   %
          \hfil $\,#$ &         %
        ~ \hfil $#$ &\vrule#&   %
          \hfil $\,#$ &         %
        ~ \hfil $#$ &\vrule#&   %
          \hfil $\,#$ &         %
        ~ \hfil $#$ &\vrule#    %
          \cr
     \noalign{\hrule}
%
%
     \noalign{\hrule}
 & \omit ~$n$
       &&  \omit \hfil $\hbox{Re}\, \lambda_n$ \hfil & \omit \hfil $\hbox{Im}\, \lambda_n$ \hfil & 
\cr 
%
     \noalign{\hrule}
    &   1    &&  \pm  3.119452    &  - 2.746676 &
\cr
    &     2  &&   \pm   5.169521     &   - 4.763570  &
\cr
    &     3  &&   \pm  7.187931      &   - 6.769565  &
\cr
    &     4  &&   \pm  9.197199     &  - 8.772481  &
\cr
    &     5  &&    \pm 11.202676     &   - 10.774162 &
\cr
    &     6  &&    \pm  13.206247    &   - 12.775239  &
\cr
    &     7  &&    \pm  15.208736   &  - 14.775979  &
\cr
    &     8  &&    \pm 17.210558    &   - 16.776515  &
\cr
    &     9  &&    \pm 19.211943    &   - 18.776919  &
\cr
    &     10  &&    \pm  21.213025    &   - 20.777232  &
\cr
    &     11  &&    \pm 23.213896  &   -  22.777489 &
\cr
    &     12  &&    \pm 25.213896  &   -  24.777489 &
\cr
    &     13  &&    \pm 27.213896 &   -  26.777489 &
\cr
    &     14  &&    \pm 29.213896 &   - 28.777489  &
\cr
    &     15  &&    \pm 31.213896  &   - 30.777489  &
\cr
     \noalign{\hrule}
                                                    }}}
\smallskip
Table 1: The lowest quasinormal 
frequencies $\lambda_{n}$ for $q=0$. 
\endinsert

\ifig\frequ{The lowest 15 quasinormal frequencies
 in the complex $\lambda$-plane 
for $q=0$.}
{\epsfxsize=9.5cm \epsfysize=5.5cm \epsfbox{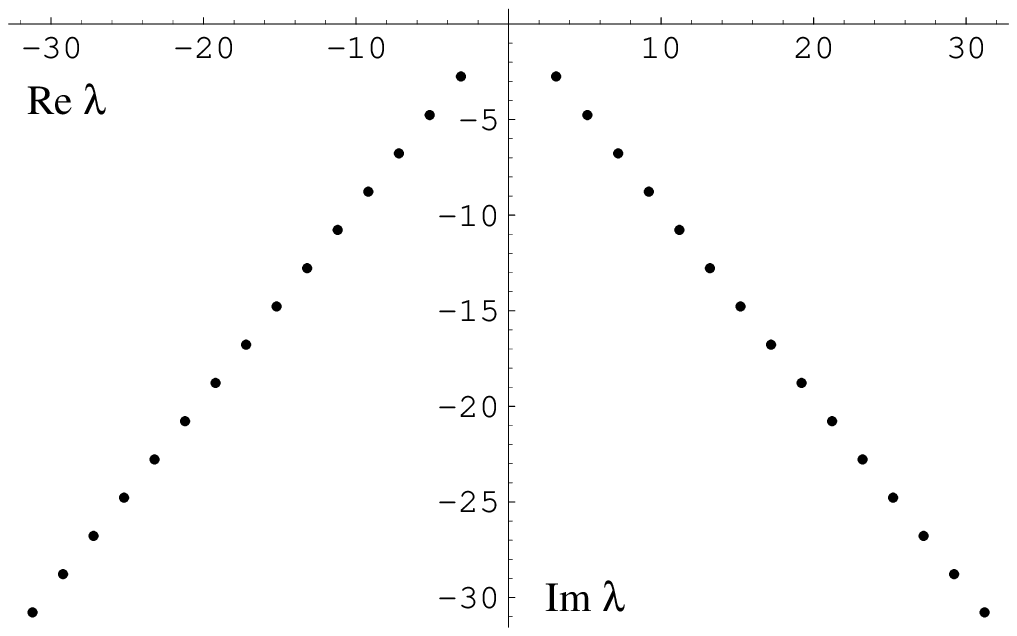}}

\ifig\rel{Re $\lam$ vs $q$ (with the interval $\Delta q = 0.5$) 
for the lowest five quasinormal frequencies. The 
zeros approach the line Re $\lam$ = $q$ as $q \rightarrow \infty$.}
{\epsfxsize=9.5cm \epsfysize=5.5cm \epsfbox{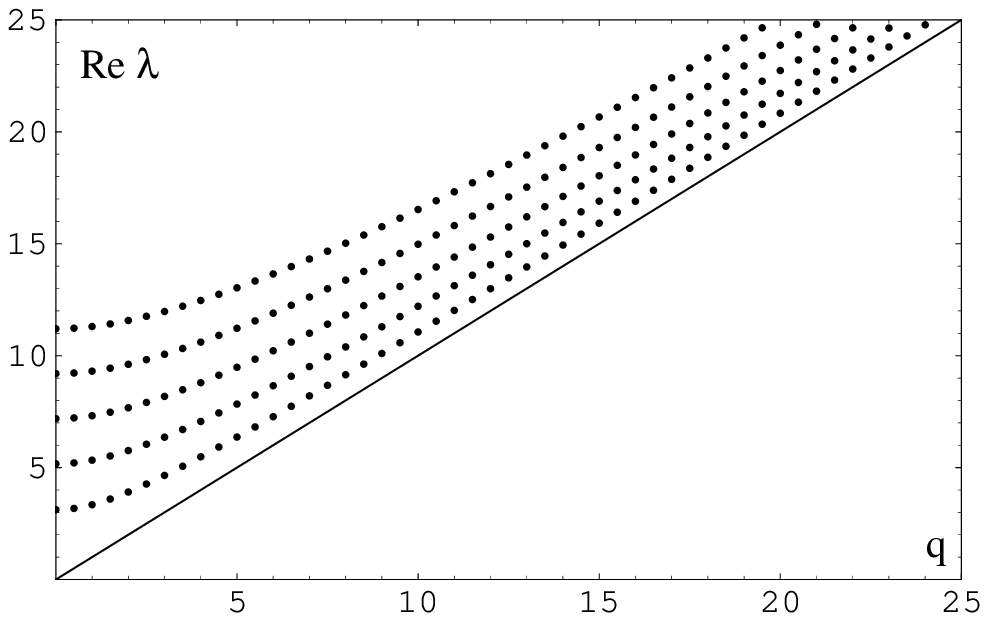}}

\ifig\imadd{Re $\lam$ vs $q$ (with the interval $\Delta q = 1$)
 for the lowest five quasinormal frequencies.}
{\epsfxsize=9.5cm \epsfysize=5.5cm \epsfbox{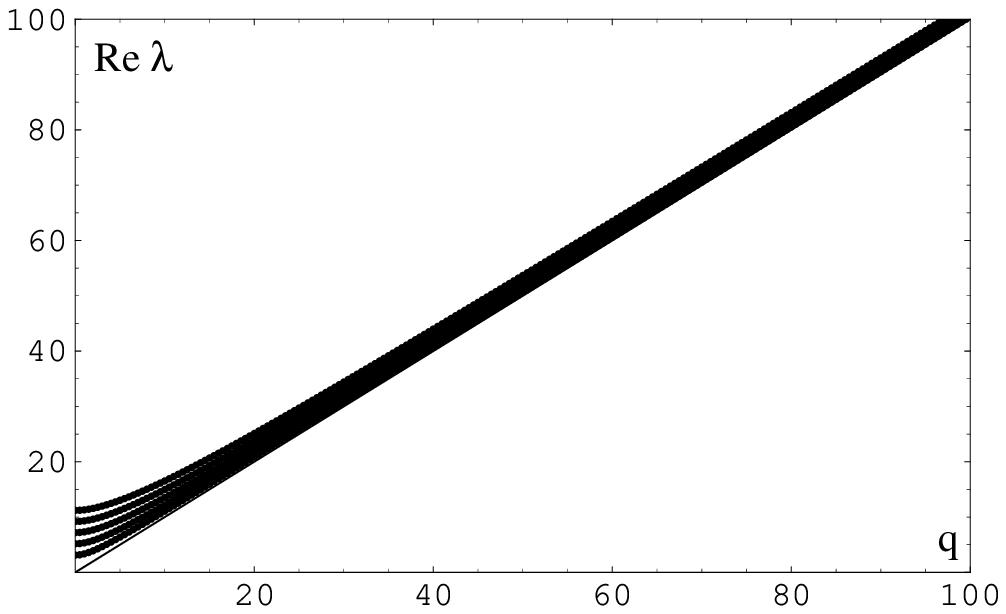}}

\ifig\iml{- Im $\lam$ vs $q$ (with the interval $\Delta q = 1$) 
for the lowest five quasinormal frequencies.}
{\epsfxsize=9.5cm \epsfysize=5.5cm \epsfbox{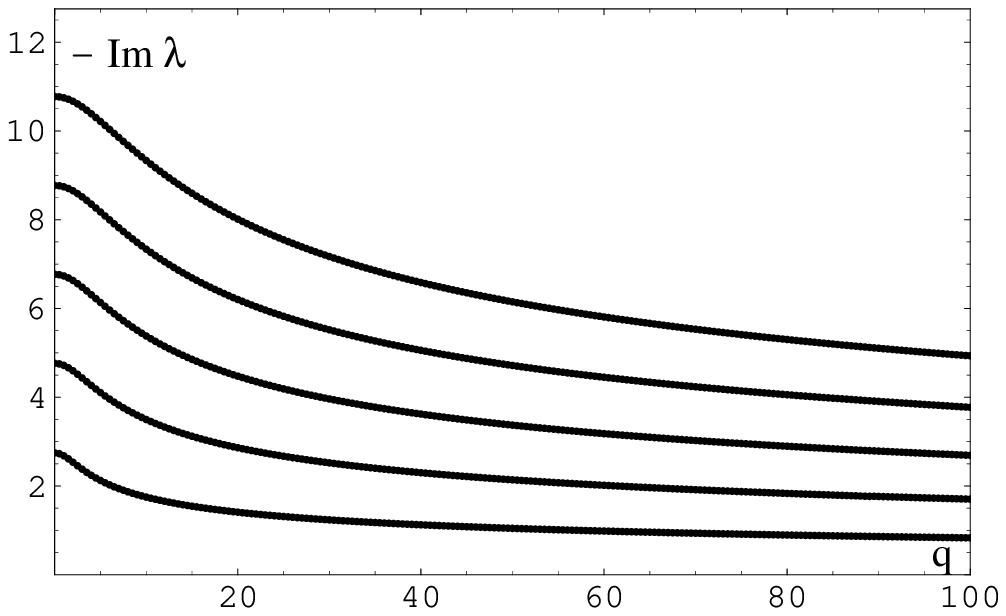}}

The lowest eigenfrequency in Table 1 can be compared with the 
result of Horowitz and Hubeny \HorowitzJD\ for a large $5d$ 
Schwarzschild- AdS black hole. Normalizing the $r_+=100$ entry 
in Table 1 of  \HorowitzJD\ appropriately ($\lam = \omega R^2/r_+$, $R=1$),
we have $\lam^{(100)} \approx 3.119627 - 2.746655\,i$ which is fairly 
close to our result $\lam^{(\infty)} \approx 3.119452 - 2.746667\,i$.

The dependence of the lowest five quasinormal frequencies on $q$ for 
$q \in [0,100]$ is
shown in Figs.~2-4. All five branches stay above the line Re $\lam = q$
(Re $\omega = |{\bf k}|$) slowly approaching it in the low-temperature 
limit $q\rightarrow \infty$. The imaginary part of the branches tends to 
zero in the same limit.  Figures 2-4 bear resemblance to the dispersion law
of thermal excitations in a weakly coupled Yang-Mills plasma\foot{Note, 
however, that in this paper we consider correlators of the gauge-invariant 
operators rather than the gluon propagator.}.

All quasinormal frequencies have negative
  imaginary parts. This fact can 
be proven rigorously using an argument based on the 
``energy-type'' integral \chandra\
or by writing Eq.~\scal\ in the Regge-Wheeler form and 
essentially repeating the proofs for AdS black holes  
given recently in \HorowitzJD , \CardosoBB . It reflects 
the stability of the near-extremal 
metric against a scalar perturbation. On the field theory side, the 
negative sign corresponds to the damping of the plasma excitations.

The computed frequencies appear to be symmetric with respect to the imaginary 
$\lam$ axis, i.e., the eigenvalue equation 
\eigen\ seems to give complex-conjugate
pairs of solutions in terms of the variable $i\lam$. This symmetry 
does not seem to be explicit in Eq.~\eigen\ \foot{The symmetry, however, is 
obvious in the original form of the equation \scal\ . I thank 
A.~V.~Shchepetilov
 for pointing this out to me.}.

\newsec{Discussion of results}

The results are compatible with expectations outlined in the 
Introduction. The retarded Green's function  has infinitely many poles in
the complex $\omega$-plane whose location depends on the spatial momentum.
There is a ``mass gap'': for small enough values of $\lam$ the propagator 
$G^R(\omega, {\bf k})$ is analytic in agreement with Eq.~\hight .
In the low-temperature limit $\lam \rightarrow \infty$, $q\rightarrow \infty$
singularities merge, forming branch cuts of the zero-temperature propagator
\foot{The zero-temperature limit \zerot\ of $G^R(\omega, {\bf k})$ can
be obtained from the Heun equation using the Langer-Olver  asymptotic expansion
(see \SonSD ).}
\zerot\ as in the BTZ case.

The distribution of quasinormal frequencies in the complex
$\omega$ plane for an asymptotically AdS background 
appears to follow a much simpler pattern than the 
one corresponding to the asymptotically flat case (compare Fig.~1 in
\leaver \ ). 
In particular, Chandrasekhar's ``algebraically special'' solution
 is absent: there are no frequencies with Re $\lam =0$.
Obviously, all statements about the behaviour of higher order modes 
are conjectural. 
It would be very desirable
 to confirm the asymptotic formula \conj\ analytically,
possibly by using the complex WKB method. We remark, however, that the 
analogous problem remains unsolved even in the much studied case of a 
Schwarzschild black hole in flat space.

Spectra of excitations in more realistic theories can be similarly studied
provided their gravitational duals are known explicitly.
It would also be interesting to determine the poles of the current and 
energy-momentum tensor correlators at finite temperature, whose 
hydrodynamic limit has recently been computed in \PolicastroSE\ .

Finally, studying gravitational quasinormal modes may be important in
investigating the stability of the non extremal black brane background.

\vskip 1.5 cm
\centerline{\bf Acknowledgments}

\vskip .2 cm
It is a pleasure to thank P.K.~Kovtun, D.T.~Son and L.G.~Yaffe 
 for discussions. I would also like to thank
A.~Nu\~{n}es-Nikitin and A.~V.~Shchepetilov for valuable comments on 
the manuscript.
This work is supported
in part by DOE grant No. DOE-ER-41132.


%

\listrefs
\end